# Electrostatic Potential of Phase Boundary in Coulomb Systems[1]

I. Iosilevski and A. Chigvintsev

*Moscow Institute of Physics and Technology*
(State University)

**Abstract.** Any interface boundary in an equilibrium system of Coulomb particles is accompanied by the existence of a finite difference in the average electrostatic potential through this boundary. This interface potential drop is a thermodynamic quantity. It depends on temperature only and does not depend on surface properties. The zero-temperature limit of this drop (along coexistence curve) is an individual substance coefficient. The drop tends to zero at the critical point of the gas-liquid phase transition. A special critical exponent can be defined to describe this behavior. The value of the discussed potential drop is directly calculated by numerical simulation of phase transitions in Coulomb systems.

Properties of the interface potential drop are discussed for several simplified Coulomb models (melting and evaporation in the One Component Plasma (OCP)). Some examples of phase transition in real situations are also discussed.

## 1. GENERAL PROPERTIES

A remarkable feature of any Coulomb system is the existence of two versions of chemical potential. The ordinary chemical potential, $\mu_i(n,T)$, is a local parameter depending on density and temperature. The electro-chemical potential, $\tilde{\mu}_i$, is the sum of ordinary chemical potential, $\mu_i(n,T)$, and average electrostatic potential, $\varphi(r)$, $\tilde{\mu}_i = \mu_i + Z_i e \varphi$. It should be stressed that the electro-chemical potential, $\tilde{\mu}_i$, is not a local parameter.

For each charged species in a Coulomb systems at equilibrium, the values of its ordinary chemical potentials in coexisting phases, $\mu_i'$ and $\mu_i''$, must not be equal under conditions of phase equilibrium. It is namely the electro-chemical potentials, $\tilde{\mu}_i$ which have the same values in coexisting phases, $(\tilde{\mu}_i)' = (\tilde{\mu}_i)''$. This equality combined with the electroneutrality condition in both phases leads to existence of the finite gap in the average electrostatic potential through the phase interface, $\Delta\varphi$.

$$\Delta\varphi \equiv \varphi''(r = +\infty) - \varphi'(r = -\infty) = [\mu_e'' - \mu_e']\, e^{-1} = [\mu_i' - \mu_i''](Ze)^{-1} \tag{1}$$

It is the non-symmetry in equilibrium properties of various charged species in condensed and gaseous phases that manifest itself by the existence of the finite gap $\Delta\varphi$. It equals to zero identically for symmetrical systems like the electron-positron plasma, the primitive ionic model of electrolyte solution, etc.

The potential drop, $\Delta\varphi = \Delta\varphi(T)$, is a thermodynamic quantity which depends on temperature only and does not depend on surface properties. The zero-temperature limit of this drop (along the coexistence curve) is an individual substance thermo-electrophysical coefficient. Its value is a supplement to the set of usual constants of the real material such as sublimation energy, ionization potential, etc.

The potential drop, $\Delta\varphi(T)$ tends to zero at the critical point of gas-liquid phase transition. A special critical exponent can be defined to describe the behavior of $\Delta\varphi(T)$ in the vicinity of the critical point.

$$\Delta\varphi(T) \sim |T - T_C|^\phi \tag{2}$$

The value of the potential drop can be directly calculated by numerical MC or MD-simulation of phase transitions in Coulomb systems when both the coexisting phases are explicitly simulated in combination.

---

[1] *Journal de Physiquie IV,* **10** Pr5 451 (2000) ("*Strongly Coupled Coulomb Systems*", Eds. C.Deutsch, B. Jankovici and M.-M. Gombert, EDP Science, Paris).

## 2. ILLUSTRATIONS

### 2.1. One Component Plasma - OCP(r)

The model is studied carefully nowadays in the version with a rigid, (non-compressible) background [1,2,7]. (Notation "r" stresses this property of background). It is known that the only phase transition in OCP(r) – Wigner crystallization – occurs in the model without any density gap ($n^*_{Fluid} \equiv n^*_{Crystal}$). Phase equilibrium condition in this case corresponds to the equality of Helmholtz free energy in both coexisting phases (notation «*» below). It should be stressed that the values of the ordinary (local) chemical potential in both coexisting phases are *not equal* in general case. Equation (1) of present work corresponds to the statement that the double electrical layer ("surface dipole") must appear at crystal-fluid interface as a result of this inequality, so that the potential of this crystal-fluid interface should compensate exactly the mean-phase deviation, $\Delta\mu^* \equiv (\mu^*_{Crystal} - \mu^*_{Fluid})$ in the ordinary chemical potential. The values of electrochemical potential in both phases will be equal in this case (1,3).

$$F(N,V,T)^*_{Crystal} = F(N,V,T)^*_{Fluid} \qquad \mu^*_{Crystal} \neq \mu^*_{Fluid} \; (!) \qquad [\tilde{\mu}_i]^*_{Crystal} = [\tilde{\mu}_i]^*_{Fluid} \qquad (3)$$

It is well known that the melting curve in OCP(r) consists of three parts [3-9] (for example, Fig.1 in [6]):
1) Low density melting of non-degenerated, classical ions ≈ line $\Gamma \equiv (Ze)^2/kTa$ = const; $\{a \equiv (3/4\pi n)^{1/3}\}$
2) High density (quantum) melting of highly degenerated ions ≈ line $r_S \equiv a/a_B \cong$ const;
3) Transition between two these parts, including the point of maximal melting temperature, $T_{max}^{**}$

<u>2.1.1. Classical Melting in OCP(r)</u> – ($\theta^{(i)} \equiv k_BT/\varepsilon_F^{(i)} \gg 1$) ↔ $\Gamma = \Gamma_m \cong 178$ [7]
In this case $\mu(\Gamma_m)_{Crystal} < \mu(\Gamma_m)_{Fluid} < 0$. Therefore crystal is *positive* and fluid is *negative* in crystal-fluid interface and we have

$$Ze\, \Delta\varphi_{melting} = [\mu_i''(\Gamma_m) - \mu_i'(\Gamma_m)] = k_BT\, (\Delta S/3Nk_B)_{melting} \cong 0.27.. \, k_BT \qquad \{(\Delta S^*/k_B)_{melting} \approx 0,82\, [8]\} \qquad (4)$$

<u>2.1.2. "Cold" (Quantum) Melting in OCP(r)</u> – ($\theta^{(i)} \equiv k_BT/\varepsilon_F^{(i)} \ll 1$) ↔ $r_S = (r_S)_m \cong 100$ [9]
In this case $0 > \mu\{(r_S)_m\}_{Crystal} > \mu\{(r_S)_m\}_{Fluid}$, so that fluid is *positive* and crystal is *negative* in crystal-fluid interface. Using results of Ceperley and Alder [9] we estimated roughly the value $(\Delta\mu)_{melting}$ at $T \to 0$.

$$(\Delta\varphi)_{melting} = (\Delta\mu)_{melting}/(Ze) \approx -0.2 \text{ V} \qquad (T \to 0) \qquad (5)$$

<u>2.1.3. Maximum Melting Temperature in OCP(r)</u> ↔ $\{T^{**} \equiv \max(T_{melt}) \approx (3 \div 10)\, 10^{-5}$ Ry [3-6]$\}$.
In this point ($T^{**}, n^{**}$) crystal-fluid equilibrium corresponds to the following special conditions:

$$(F^{**})_{Crystal} = (F^{**})_{Fluid} \qquad (P^{**})_{Crystal} = (P^{**})_{Fluid} \qquad (\mu^{**})_{Crystal} = (\mu^{**})_{Fluid} \qquad (6)$$

Therefore one obtains for discussed potential drop of crystal-fluid interface in point ($T^{**}, n^{**}$)

$$(\Delta\varphi^{**})_{melting} = 0 \qquad (7)$$

Consolidation of all three parts of total dependence $\Delta\varphi_{melting}(T)$ is exposed at general picture (Figure 1).

### 2.2. One Component Plasma - OCP(c)

This version of OCP with uniform *and* compressible background (it is stressed by notation "c") was considered earlier in [8] and has been studied more carefully in [10-13]. Compressibility of *uniform* background leads to the two sequences: appearance of the finite density gap at melting [8] and to the new first-order phase transition of gas-liquid type with upper critical point [10-13]. Phase equilibrium conditions in OCP(c) have standard form of the conditions in ordinary materials:

$$[P^{(+)} + P^{(-)}]_{Crystal} = [P^{(+)} + P^{(-)}]_{Fluid} \qquad [\mu^{(+)} + Z\mu^{(-)}]_{Crystal} = [\mu^{(+)} + Z\mu^{(-)}]_{Fluid} \qquad (8)$$

$$[n^{(+)} + Zn^{(-)}]_{Crystal} = [n^{(+)} + Zn^{(-)}]_{Fluid} = 0$$

Here $P^{(-)}$ and $\mu^{(-)}$ are the thermodynamic contributions of background (for example, of an ideal electrons gas). Note that the values of ionic chemical potential in coexisting phases of OCP(c) are still not equal.

$$[\mu^{(+)}]_{Fluid} \neq [\mu^{(+)}]_{Crystal} \qquad (9)$$

Existence of corresponding potential drop, $\Delta\varphi(T) = [\mu^{(+)}]_{Fluid} - [\mu^{(+)}]_{Crystal}$, through the phase boundary in OCP(c) was claimed and calculated in [10] and studied in details in [11] for the variants of OCP(c). It can be proved or directly calculated that the value of $\Delta\varphi(T)$ in OCP(c) is higher then that in OCP(r). General picture for $\Delta\varphi_{melting}(T)$ in OCP(r) and OCP(c) is exposed at Figure 1.

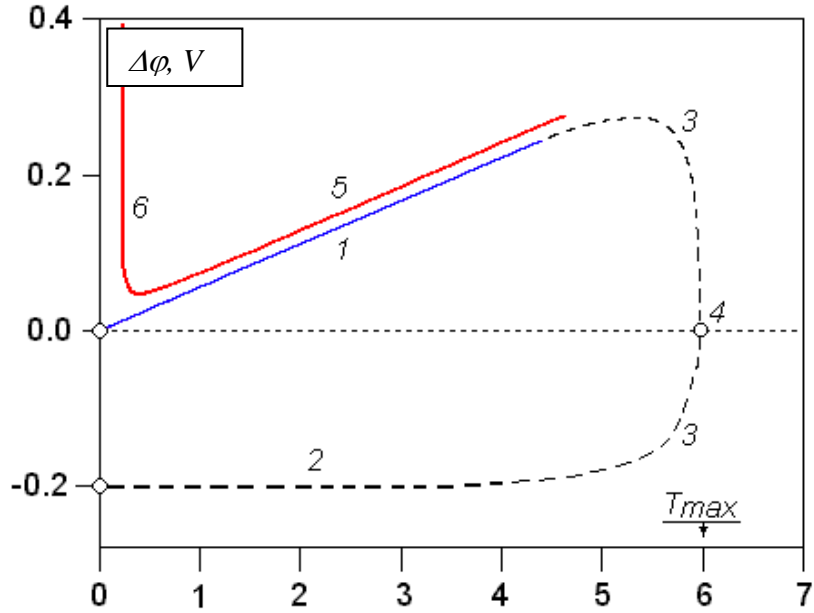

**Figure 1**. Potential of crystal-fluid interface in OCP. {$1 \div 4$ – OCP(r); $5,6$ – OCP(c)}
*1* – Melting of non-degenerated, classical ions in OCP(r) {$\Gamma \equiv (Ze)^2/kTa \cong 178$ [7]}; *2* – Quantum melting of highly degenerated ions in OCP(r) {$r_S \equiv a/a_B \cong 100$ [9]}; *3* – Transition between *1* and *2*; *4* – $T_{max}^{**}$ – maximal melting temperature according to [6]; *5* – Melting of non-degenerated, classical ions in OCP(c) of ions ($Z = 2$) on background of ideal gas of degenerated electrons; *6* – The same as *5* in low density limit (weakly degenerated *uniform* background). (*1,5* and *6* – present calculation; *2* – estimation from results [9]; *3* – qualitative picture)

Potential drop $\Delta\varphi(T)$ tends to zero at critical point of gas-liquid phase transition. A special critical exponent can be defined to describe this behavior of $\Delta\varphi(T)$ in the vicinity of critical point.

$$\Delta\varphi(T) \sim |T - T_C|^\phi \qquad (10)$$

This exponent, $\phi$, coincides with the density exponent, $\beta$, if chemical potential of each component, ions and electrons, are analytic functions of temperature and densities at critical point. It is really so for the *model* OCP(c) {$\phi = \beta = 1/2$}

**2.3. Gas-Liquid Coexistence in Simple Metal ($Z = 1$)**

We consider coexistence of electron-ionic system in condensed phase, $n_i' + n_e'$, with electron-ion-atomic system in vapor phase, $n_i'' + n_e'' + n_a''$. Equilibrium conditions include equality of temperature, pressure and Gibbs free energy in both phases. These equations (11,12) are supplemented with equation of ionization equilibrium in vapor (13) and with electroneutrality conditions for both phases (14)

$$P'(T, n_i', n_e') = P''(T, n_i'', n_e'', n_a'') \tag{11}$$

$$\mu_i' + \mu_e' = \mu_i'' + \mu_e'' \tag{12}$$

$$\mu_i'' + \mu_e'' = \mu_a'' \tag{13}$$

$$n_i' = n_e' \qquad n_i'' = n_e'' \tag{14 a,b}$$

The equations (11-14) fix all the concentrations: $n_i'$, $n_e'$, $n_i''$, $n_e''$, $n_a''$. It should be stressed that in general case the values of local chemical potential of electrons (as well as for ions) are not equal in both phases.

$$\mu_i' \neq \mu_i'' \qquad \mu_e' \neq \mu_e'' \tag{15 a,b}$$

It is namely the electro-chemical potentials, $\tilde{\mu}_i$ which have the same values in coexisting phases (16). This condition fixes the interface potential drop, $\Delta\varphi(T)$ (1)

$$(\tilde{\mu}_i)' = \mu_i' + Ze\varphi' = \mu_i'' + Ze\varphi'' = (\tilde{\mu}_i)'' \tag{16 a}$$

$$(\tilde{\mu}_e)' = \mu_e' - e\varphi' = \mu_e'' - e\varphi'' = (\tilde{\mu}_e)'' \tag{16 b}$$

Simple relation for $\Delta\varphi(T)$ may be obtained in the limit $T \to 0$ along the coexistence curve. In this limit $\mu_a' \to \mu_a^0 = const$. The vapor phase is ideal and $\Delta\mu_{i,e}'' \equiv \mu_i'' - \mu_e'' \to 0$. It leads to the equation:

$$e\Delta\varphi(0) = [(\Delta_s H^0 + I)/2 - \{\mu_e(0)\}_{Cond}] \tag{17}$$

Here $\Delta_s H^0$ is the sublimation energy of metal at $T = 0$; $I$ – atomic ionization potential and $\{\mu_e(0)\}_{Cond}$ – zero-temperature limit of electronic chemical potential in condensed phase.

## Acknowledgements

The work was supported by "Universities of Russia", Grant 2550.